\documentclass{article}[10pt]

\usepackage{jucs2e}
\usepackage{url}

\usepackage{epsfig}
\begin{document}

\title{Experiments in Clustering Homogeneous XML Documents to Validate
an Existing Typology}

\author{
{\bfseries Thierry Despeyroux}\\
{\bfseries Yves Lechevallier}\\
{\bfseries Brigitte Trousse}\\
{\bfseries Anne-Marie Vercoustre}\\
   (AxIS research team, Inria, France\\
   FirstName.LastName@inria.fr)\\
}

\maketitle

\begin{abstract}
This paper presents some experiments in clustering homogeneous XML
documents to validate an existing classification or more generally an
organisational structure.
Our approach integrates techniques for extracting knowledge from
documents with unsupervised classification (clustering) of documents.  
We focus on the
feature selection used for representing documents and its impact on the emerging
classification.  We mix the selection of structured features with fine textual selection based on syntactic characteristics.
We illustrate and evaluate this approach with a collection of 
Inria activity reports for the year 2003. The
objective is to cluster projects into larger groups (Themes), based on
the keywords or different chapters of these activity reports.  We then
compare the results of clustering using different feature selections,
with the official theme structure used by Inria.

\end{abstract}


\begin{category}
H.3.1, I.5.3, I.5.4
\end{category}

\section{Introduction}

With the increasing amount of available information,
sophisticated tools for supporting users in finding useful
information are needed. In addition to tools for retrieving relevant documents,
there is a need for tools that synthesise and exhibit information that
is not explicitly contained in the document collection, using document
mining techniques.  Document mining objectives include extracting
structured information from rough text, as well as document
classification and clustering.

Classification aims at associating documents to one or several
predefined categories, while the objective of clustering is to
identify emerging classes that are not known in advance.  Traditional
approaches for document classification and clustering rely on various
statistical models, and representation of documents mostly based on
bags of words. An important characteristic of text categorisation is the size
of the vocabulary, which is often referred as the high dimension of
the feature space. Automatic feature selection methods have been
proposed to reduce the dimension of the space. They usually try to
identify representative words that can discriminate documents between
various classes. For a comparison of those methods for classification
see \cite{yang}. 

XML documents are becoming ubiquitous because of their rich and
flexible format that can be used for a variety of applications.
Standard methods have been used to classify XML documents, reducing
them to their textual parts. These approaches do not take advantage of
the structure of XML documents that also carries important information.

In this paper we focus on XML documents and we study the impact of
selecting (different) parts of the documents for a specific clustering
task. The idea is that different parts of XML documents correspond to
different dimensions of the collection that may play different roles
in the classification task. We therefore consider two levels of
feature selection: 1)  coarse selection at the structure level and 2)  fine linguistic selection of words within the text of elements.

Based on the selected features the documents are then clustered using
a dynamical classification algorithm that builds a prototype of each
cluster as the union of all the features (words) of the documents
belonging to this cluster. Furthermore, for each resulting cluster, we
can exhibit the words that discriminate this cluster.

Our experimentations use the collection of activity reports that
were produced by the research groups at Inria. The task
is to identify meaningful themes that would group projects working in
related research domains.  These groups are then compared to two
Expert grouping, the first one used by Inria until 2003, and the new
one proposed in 2004.
 
\section{Inria Activity Report}

Every year, Inria (The French National Institute for Research in
Computer Science and Control) publishes an activity report (RA) 
made available to the French parliament and to our
industrial and research partners. Traditionally produced as a paper
document, this report is now published as a CD-Rom and the scientific
part is made available on the Web since 1996 (in HTML and PDF). It is a
collection of reports written by every Inria research team (in English since 2002). The XML version of these documents contains 139 files, 
a total of 229 000 lines, more than 14.8 Mbytes of data.

If the logical structure is defined by a DTD, the overall style and content are very flexible and unconstrained. The top level part of the DTD is given below:
\vspace{-0.2 cm}
\begin{verbatim}
<!ELEMENT raweb (header, moreinfo?, members, presentation,
              foundation?, domain?, software?,  results,
              contracts?, international?, dissemination?, biblio)>
<!ATTLIST raweb year CDATA #IMPLIED >
\end{verbatim}
\vspace{-0.2 cm}

Mandatory sections include the list of team members, {\it presentation} of objectives, new {\it results}, and the
list of publications for the year (biblio). Optional sections include research {\it foundation}, application domains, software, as well as international and national cooperations.

Inria research teams are also grouped into scientific {\it themes} that act as virtual structures for the purpose of
presentation, communication and evaluation. The number of themes and
allocation of teams to themes were decided some years ago by the
board of directors and have changed recently. Choice of themes
and team allocation are mostly related to strategic objectives and
scientific closeness between existing teams.

This has motivated our experiments in comparing the automatic
clustering of teams, based on self-descriptions in their activity
reports, with the two sets of themes defined by the Organisation. We
will call them Expert Themes 2003 and 2004 respectively.
Without anticipating on the results of the experiments, we are
interested in discovering possible natural grouping of teams,
identifying the keywords that better characterise those groups, and
the potential difference with the organisational structures. 

%

\section{Methodology for Clustering XML Documents}

As said above, our objective is to cluster the research teams in
themes, using their activity reports as data source. We hypothesis that activity reports reflect the research domains the teams are
involved in and that some parts or the reports are more representative than others in describing research. For example, conferences and journals where
researchers publish are indicative of their research fields.

\subsection{Structure Selection and Vocabulary Definition}

The first step consists in selecting various parts of the XML documents that may be relevant for the classification task.  
This extraction uses the tools described in \cite{despeyroux} to extract the text of elements, but standard XML tools could be used instead when the extraction does not require any inference.

As we expect that various parts of the activity report would play different roles in classifying teams, we ran five experiments using different parts of the activity report, that are well-identified XML elements. We call this process
"structured feature selection". In this step, the documents are represented by the text of the selected elements.

\begin{enumerate}

\item Experiment  K-F: Keywords attached to the {\it foundation} part
\vspace{-0.3 cm}
\item Experiment  K-all: Keywords, whatever the sections they are
attached to.
\vspace{-0.3 cm}
\item Experiment T-P: Full text of the {\it presentation} part
\vspace{-0.3 cm}
\item Experiment T-PF: Full text of the {\it presentation} and 
{\it foundation} parts
\vspace{-0.3 cm}
\item Experiment T-C: Names of conferences, workshops, congress, etc,
\end{enumerate}

The goal of these experiments is to evaluate which parts are more relevant for the clustering task.

The second processing step consists in the automatic selection of significant words within the previously selected texts. Classical methods of textual feature selection are based on statistical approaches, for example selection based on word frequency (DF) or information gain (IG). These methods works well for large collections of texts and involved pre-processing of the full collection. In our case the frequency of words may vary depending on the selected parts of documents and the size of the resulting collection can be very heterogeneous from one experiment to the other.  In order to avoid heterogeneous frequency, we chose a natural language approach where words are tagged and selected according to their syntactic role in the sentence. We use TreeTagger, a tool for annotating text with part-of-speech and lemma information, developed at the
Institute for Computational Linguistics of the University of Stuttgart
\cite{schmid,schmida}.

We retain
different types of words, depending on the structured feature
selection. For experiments K-F and K-all (keywords) we keep nouns, verbs,
adjectives, (excluding conjunctions, unknown words, etc.), while for
experiments T-PF and T-P (full text), we keep only the nouns to limit the
number of features.
There is one difficulty with conference names due to their very free
and heterogeneous labelling : some teams would use the full name of the
conference, others the acronym in various formats (e.g. POPL'03,
POPL03, POPL 2003). We therefore built manually a normalized list of
all the conference names and matched automatically the form used in
the RA with the normalized form. Since conference acronyms are significant and unknown to the tagger, we decided not to use the tagger for this experiment, keeping all the words but stop words (such as proceedings, conference, etc.).

Finally, for all experiments, words that are not used at least by two
teams are removed.  Table \ref{table:datasize} summarises the size of data (words) used in each experiment. 

\begin{table}
\begin{center}
{\scriptsize
\begin{tabular}{|l|r|r|r|r|}\hline
 Experiences  & number   &  extrated   & selected   &  voca-  \\
        & of projects   &  words   &  words  &  bulary  \\ \hline
K-F & 80 & 2234  &  1053 & 134 \\
K-all & 121 & 8671 &   6171 & 382 \\
T-P & 138 & 63711 &  16036 & 365 \\
T-PF & 139 & 320501 &  87416 & 809 \\
T-C & 131 & 10806 & 7915 & 887\\  
 \hline
\end{tabular}} \caption{\label{table:datasize}Size of data for the various experiments}
\end{center}
\vspace{-0.5 cm} 
\end{table}
\subsection{Clustering Method and External Evaluation}
The third step is clustering of documents in a set of disjoint classes using the vocabulary defined for the five experiments as described above.

Our clustering algorithm is based on the partitioning method proposed by 
\cite{celeux}, where the distances between clusters is based on the frequency of the words of the selected vocabulary. This approach is equivalent to the k-means algorithm. As for the k-means we  
represent the clusters by
prototypes which summarize the whole information of the 
document's belonging to each of them.
 
More precisely, if the vocabulary counts $p$ words, each document $s$ is represented by the vector 
$x_s=(x_s^1,...,x_s^j,...,x_s^p)$ where $x_s^j$ 
is the number of occurences of word $x_j$ in the document $s$, then the prototype $g$ for a class $U_i$ is represented by
$g_i=(g_i^1,...,g_i^j,...,g_i^p)$ with 
$g_i^j=\sum_{s \in U_i}{x_s^j}$.

Finally, the prototype of each class been fixed, every element is assigned to a class according to its proximity to the prototype. The proximity is
measured by  a classical distance between distributions (e.g. chi-squared).

We evaluate the quality of our automatic clustering by
comparing the results with the two sets of themes used by Inria. We
call this evaluation {\it external validity}, since the clustering process 
does not involve those themes. For all quantitative evaluations
we use two complementary measures: the {\it F-measure} and the {\it corrected Rand
index}.

The {\bf F-measure} proposed by \cite{jardine} combines the
precision and recall measures from information retrieval and treats each cluster as if it
was the result of a query and each class as if it was the desired
answer to that query. For a priori group $U_i$; and cluster $C_j$, recall(i,j) is equal to $n_{ij}$/$n_{i.}$ and 
precision(i,j) is equal to $n_{ij}$/$n_{.j}$,
where $n_{ij}$
is the number of documents in a group $U_i$ and
the cluster $C_j$; $n_{i.}$ the number of documents in a
priori group $U_i$; $n_{.j}$ the number of documents in
cluster $C_j$.
Then, the F-measure between $U_i$ and $C_j$ is
given by
F(i,j)=(2.*recall(i,j)*precision(i,j)/(recall(i,j)+precision(i,j)).

The F-measure between a priori partition $U$ and the partition $C$ in K clusters
is given by:
\begin{equation} \label{eq:F}
F=\sum_{j=1}^{k}{\frac{n_{.j}}{n}*\max_j (F(i,j))}
\end{equation}, where $n$ is the total number of documents in the data set.

The  {\bf corrected Rand} ($CR$) index is
defined in \cite{hubert} for comparing two
partitions. We remind its definition. Let $U = \{U_1, \dots, U_i, \dots, U_r\}$ and $P = \{C_1, \dots,
C_j, \dots, C_k\}$ be two partitions of the same data set having
respectively $r$ and $k$ clusters. The corrected Rand index is:

\begin{equation} \label{eq:adjustedrand} CR =
\frac{\sum_{i=1}^r\sum_{j=1}^k {{n_{ij}}\choose{2}} -
{{n}\choose{2}}^{-1}\sum_{i=1}^r{{n_{i.}}\choose{2}}\sum_{j=1}^k{{n_{.j}}\choose{2}}}
{\frac{1}{2}[\sum_{i=1}^r{{n_{i.}}\choose{2}}+\sum_{j=1}^k{{n_{.j}}\choose{2}}]
-{{n}\choose{2}}^{-1}\sum_{i=1}^r{{n_{i.}}\choose{2}}\sum_{j=1}^k{{n_{.j}}\choose{2}}}
\end{equation}

\noindent where ${{n}\choose{2}}=\frac{n(n-1)}{2}$ and $n_{ij}$, $n_{i.}$, $n_{.j}$ and $n$ are defined as above.\\

To conclude, the F-measure is easier to interpret and can support local analysis
(through the $F_{ij}$), while the Rand gives a measure of the
significance of the results for a given number of clusters.

\subsection{Results Analysis}
Table \ref{table:externalvalid} summarizes our results for different feature 
selections and different number of clusters (4, 5 and 9). We first analyse 
results for Themes 2003 and Themes 2004 separately, then compare between 
the two.

For Themes 2003, we get the best results consistently for the two measures
and all features when clustering into 4 clusters. One exception is 
clustering in 9 sub-themes using the text of both {\it presentation} and 
{\it foundation} (T-PF-c). The overall best result is obtained with four 
clusters using {\it presentation} and {\it foundation} (T-PF-a). 
A finer analysis using sub-themes (not presented here by lack of space), highlights good mapping
between clusters and sub-themes.

For Themes 2004, we get good results for clustering in 5 or 4 clusters, 
with the best results with 5 clusters when using all the keywords.

In both cases, the sections about {\it Foundation} seem representative 
of the research domains, either through the full text of those sections 
or through their attached keywords, for the teams who provided such 
keywords.

Overall, our automatic clustering compares better with Themes 2003 than 
with Themes 2004, with the exception of using all keywords for creating 
5 clusters.
There is not much difference when comparing with Themes2003 or 
Themes 2004 when using the conference names. Somehow disappointing 
results with conference names may be explained by not using the tagger, 
leaving us with too many different words (see table \ref{table:datasize}).

We also note that the two measures, F-measure and corrected rand, are 
consistent trough the experiments: high F-measure scores correspond to 
good rand values.

\begin{table}
{\scriptsize 
\begin{tabular}{|l|r|r|r|r|r|r|r|}\hline
 Exp.  &  Nb. of & F   & Rand   &  F   &  Rand &  F   &  Rand \\
 & clusters & Themes 2003  &  Themes 2003   &  subthemes  &  subthemes & Themes 2004  &  Themes 2004\\ \hline
K-F-a   & 4 & {\bf 0.53} & {\bf 0.14} & 0.38 & 0.09 & 0.46 & 0.11 \\
K-F-b   & 5 & 0.44 & 0.05 & 0.35 & 0.06 & 0.37 & 0.03  \\
K-F-c   & 9 & 0.42 & 0.10 & 0.37 & 0.08 & 0.43 & 0.12\\ \hline
K-all-a & 4 & 0.52 & 0.17 & 0.36 & 0.09 & 0.47 & 0.15\\
K-all-b & 5 & {\bf 0.53} & {\bf 0.17} & 0.37 & 0.10 & {\bf 0.54} & 0.22 \\
K-all-c & 9 & 0.46 & 0.13 & 0.40 & 0.12 & 0.38 & 0.10\\  \hline \hline
T-P-a   & 4 & {\bf 0.55} & 0.19 & 0.40 & 0.14 & 0.50 & 0.19 \\
T-P-b   & 5 & 0.45 & 0.11 & 0.42 & 0.12 & 0.47 & 0.15 \\
T-P-c   & 9 & 0.44 & 0.11 & 0.45 & 0.16 & 0.44 & 0.14 \\ \hline
T-PF-a  & 4 & {\bf 0.66} & {\bf 0.32} & 0.49 & 0.27 & 0.50 & 0.21\\
T-PF-b  & 5 & 0.56 & 0.22 & 0.43 & 0.18 & {\bf 0.51} & 0.20 \\
T-PF-c  & 9 & 0.48 & 0.22 & {\bf 0.55} & 0.29 & 0.46 & 0.19 \\ \hline
T-C-a   & 4 & {\bf 0.51} & 0.15 & 0.39 & 0.15 & 0.50 & 0.21 \\
T-C-b   & 5 & 0.44 & 0.18 & 0.45 & 0.24 & 0.47 & 0.17 \\
T-C-c   & 9 & 0.45 & 0.13 & 0.47 & 0.21 & 0.45 & 0.15 \\
 \hline
\end{tabular}} \caption{\label{table:externalvalid}Results by external validity}
\vspace{-0.5 cm}
\end{table}

\section{Related Work}

Currently research in classification and clustering methods
for XML or semi-structured documents is very active. 
New document models have been proposed by (\cite{yi}, \cite{denoyer}) to extend the
classical vector model and take into account both the structure and the
textual part.  It amounts to distinguish words appearing in
different types of XML elements in a generic way, while our approach
uses the structure to select (manually) the type of elements relevant
to a specific mining objective.

XML document clustering has been used mostly for
visualizing large collections of documents, for example \cite{guillaume}
cluster AML (Astronomical Markup Language)
documents based only on their links.\\
 \cite{jianwu} propose
a model similar to \cite{yi} but adding in-
and out-links to the model, and they use it for clustering rather than
classification. \cite{yoon} also propose a BitCube 
model for clustering that represents documents based on their ePaths 
(paths of text elements) and textual content.
Their focus is on evaluating time performance rather than clustering 
effectiveness.

Another direction is clustering Web documents returned as answers to a
query, an alternative to rank lists. \cite{zamir} propose
an original algorithm using a suffix tree structure, that is linear in
the size of the collection and incremental, an important feature to
support online clustering.

\cite{larsen} compare different text feature extractions,
and variants of a linear-time clustering
algorithm using random seed selection with center
adjustment.

\section{Conclusion and Future Work}

In this paper we have presented a methodology for clustering XML 
documents and evaluate the results, for different feature selections, in 
comparison with two existing typologies.
Although the analysis is closely related to our specific collection,
we believe that the approach can be used in other contexts, for 
other XML collections where some knowledge of the semantic of the DTD is 
available.

The results show that the quality of clustering strongly depends on the 
selected document features. In our application, 
clustering using {\it foundation} sections always outperforms clustering 
using {\it keywords}. This conclusion can be turned the other side down, 
as an indication for the organization that some parts of the Activity 
Report do not appropriately describe the research domains and that the 
choice of keywords and research presentation could be improved to carry 
a stronger message.

On more technical aspects, our approach provides a flexible clustering 
framework where structured features and textual features can be selected 
independently, although comparisons should be done with textual feature selection based on statistical approach (tf*idf).

Finally we plan to carry further experiences with conference names using 
an ontology of conferences. A first step would be to build a good 
classifier able to match incomplete and incorrect conference or journal 
titles with their normalized forms.

{\bf Acknowledgements}: The authors wish to thank Mihai Jurca, engineer in the 
AxIS team
for his useful support in the feature pre-processing of the data.

\end{document}